\documentclass[prb,twocolumn,showpacs]{revtex4}
\usepackage{graphicx}
\usepackage{epstopdf}
\usepackage{amsfonts}
\usepackage{amsmath}
\usepackage{amssymb}
\usepackage{bm}
\usepackage{hyperref}
\begin{document}
\title{Electronic structure and nontrivial  topological  surface states  \\
in ZrRuP and  ScPd$_{3}$ compounds}
\author{Armindo S. Cuamba$^{1}$}
\author{Hong-Yan Lu$^{2}$}
\author{Lei Hao $^{3}$}
\author{C. S. Ting$^{1}$}

\affiliation{$^1$ Texas Center for Superconductivity and Department of Physics, University of Houston,
 Houston, Texas 77204, USA\\
 $^2$ School of Physics and Electronic Information, Huaibei Normal University, Huaibei 235000,China  \\
 $^3$ Department of Physics, Southeast University, Nanjing 210096, China
 }

\date{\today}

  \begin{abstract}
 In this paper, we investigate the topological properties of ZrRuP and ScPd$_{3}$ using the first-principles calculations. We calculate the electronic structure, surface states, and Z$_{2}$ topological invariant.  The band structure of ZrRuP without the spin-orbit coupling (SOC) exhibits a nodal line located  at  $\Gamma$-K-M high symmetry line in the Brillouin zone(BZ). The nodal line is protected by the mirror symmetry corresponding to the plane $k_{z}=0$. The geometry of Fermi surface is a connected torus centered at K point with open orbits at the boundary of BZ, when the Fermi level is  shifted up 0.25 eV. The inclusion  of SOC leads to the gap opening in the band structure which corresponds adiabatically to transition from nodal line semimetal to topological insulator. The corresponding surface states were obtained from thin film calculations, and the result shows surface states emerging along the gap region. We also identify the ScPd$_{3}$  compound as a Dirac nodal line semimetal. This structure has nontrivial surface states and the corresponding Z$_{2}$ topological invariant is (1;000). Both the ScPd$_{3}$ and ZrRuP have the Fermi surfaces of the bands which make nodal lines with open orbits.
  \end{abstract}

  \maketitle

  \section{introduction}

  The investigation of material with nontrivial topology of band structure has become an exciting topic recently after discovering of interesting properties such as the conducting edge states\cite{Wang-2013} and quantum spin hall effect\cite{Konig-2007} in a topological insulator(TI). The strong spin-orbit coupling(SOC) creates  band inversion in some insulator material which can lead to nontrivial band topology. In addition, the corresponding  surface states  are protected  by the time-reversal symmetry and inversion symmetry. The TIs have been realized in  one-dimensional structure with helical edge states\cite{Javier-2017}, in the two-dimensional structure on HgTe\cite{Andrei}( which is quantum spin hall insulator\cite{Konig-2007}), and in three-dimensional structures in SnTe\cite{Tanaka-2012}. Moreover, the superconductivity was reported in the Cu doped  topological insulator Bi$_{2}$Se$_{3}$\cite{Hor-2010}.

  The realization of TI led to a search and discovery of new class of materials, the topological semimetal, and metal\cite{Heung-2015} structures.  Topological semimetals are classified based on  bands  touching  in the momentum space and are protected by certain symmetry. When the bands touch results in degenerate point node the material is  Dirac semimetal \cite{Liu-2014-Dirac,Liu-2014-Na3Bi,Liu-2014-Cd3As3,Chang-2017,Guo-2017,Le-2016,Huang-2016,Zhang-2017,Yan-2017,Han-2017}, the continuous touching of degenerate bands makes the nodal line semimetal\cite{Huang-2016-nodalline}, and  Weyl semimetal \cite{Yang-2015,Xu-2016-TaP,Sun-2015,Deng-2016,Alexey-2015,Yan-2015-MoTe2,Su-2017-LaAlGe,Koepernik-2017} corresponding to nondegenerate point node.  Recently the type-II Dirac cone was verified experimentally while the type-II  nodal line was predicted theoretically. The band crossings  in this material are characterized by the tilting of the band which leads to violation of Lorentz invariance. The nodal line  phase can exist with or without SOC,  depending on the magnitude of the SOC in the material. For instance, the CaAgP\cite{Yamakage-2016} is topological line-node semimetal with SOC, and the topological surface states are protected by the mirror symmetry, and CaAgAs\cite{Yamakage-2016} has a  nodal line without SOC and is equivalent to a topological  insulator.

  Here we study  the ZrRuP and ScPd$_{3}$  compounds. The ZrRuP  crystallizes in  hexagonal structure  and  exhibit superconductivity with the transition temperature T$_{c}$=13 K\cite{Shirotani-1993}. The electronic structure  of ZrRuP\cite{Izumi-2002,Izumi-2003} and ScPd$_{3}$\cite{Jeong-2006} have been studied. However, there is barely any investigation of topological properties of these materials. The  Fermi surfaces of the bands that make the nodal lines have open orbits which make the structure topological nontrivial. Till now there are  no studies of topological properties of these compounds. So, more investigations on these materials are required in order to determine their topological nature.

  In this paper, we report our first-principles calculations on topological properties of ZrRuP and ScPd$_{3}$ compounds, and we predict  that they are  adiabatically equivalent to  a topological insulating phase with the inclusion of SOC. We calculate the band structure and   the corresponding Fermi surfaces with and without SOC, and the topological properties. The band structure of ZrRuP without the SOC shows a  nodal line formed by the high dispersive bands located above the Fermi level.  The nodal line is located at k$_{z}$=0 plane, it makes a  close line around K point and is protected by mirror symmetry.  The Fermi surface calculated at 0.25 eV crosses the bands that make the nodal line and its geometry consists of a connected torus with open orbits. The inclusion of SOC leads to  band-gap opening across the Brillouin zone, and few  bands cross the Fermi level and the nodal line disappears. We calculate the surface states by using the tight binding Hamiltonian with  20 slabs for a thin film of ZrRuP. The surface states appear along the gap and are located between the $\Gamma$-K-M. The obtained topological invariant Z$_{2}$ is (0,001). The band structure of ScPd$_{3}$ indicates the existence of nodal line above and below the Fermi level. This structure has nontrivial topological surface states. These results can be confirmed by  future experiments, such as the angle-resolved photoemission spectroscopy (ARPES) experiments.

  \section{Computational Details and crystal structure}

\begin{figure}
  \includegraphics[width=3in]{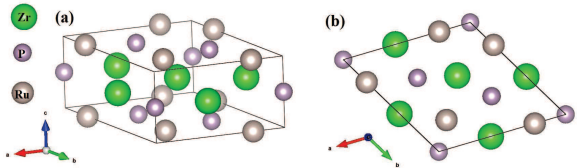}
  \caption{\label{fig:1}(colour online) Crystal structure of the ZrRuP. The crystal structure corresponds to noncentrosymmetric symmetry because it breaks the inversion symmetry.}
    \end{figure}

    Fig.\ref{fig:1} (a) and (b) shows the crystal structure of ZrRuP compound. The crystal parameters are obtained from X-rays diffraction experiments which are, $a$=$b$=6,455 \AA, and $c$=3.817 \AA\cite{Izumi-2002}. The ZrRuP compound has  non-centrosymmetric structure and crystallizes in $P-62m$ (189) space group which corresponds to the hexagonal  structure. In order to investigate the topological properties of ZrRuP and ScPd$_{3}$ compounds, we use density functional theory (DFT) which is implemented in VASP (Vienna $ab$ $initio$ simulation package)\cite{Kresse-1996}, and quantum-espresso package and wannierTools\cite{Wu-2017}.

   The 10$\times$10$\times$10 grid in the Brillouin zone is used to perform the self-consistent calculations for the electronic structure with and without spin-orbit interaction. The cut-off energy of the wave function expansion is 400 Ry. In the calculations, the ultra-soft pseudopotential is used and the exchange-correlation function potential is the generalized gradient approximation with Perdew-Burke-Ernzerhof (PBE).  The thin film calculation is obtained by using the tight binding Hamiltonian from Wannier function\cite{Sauza-2001,Marzari-1997}. The tight binding Hamiltonian is built from the Wannier function by using the $s$, $p$ and $d$ orbital for Zr, Ru, Sc, and Pd atoms and $s$, $p$ orbital for P.

  \section{Results and discussion}
  In this section we discuss  the electronic structure and topological properties of ZrRuP and ScPd$_{3}$.
 \subsection{Electronic structure of ZrRuP}

  \begin{figure}
  \includegraphics[width=3in]{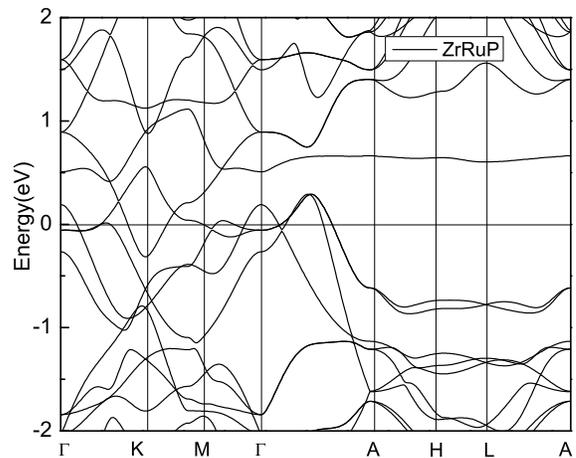}
  \caption{\label{fig:2} Band structure of ZrRuP calculated without SOC along the high symmetry lines on the hexagonal Brillouin zone. The crossing of the conduction and valence band from M-K and K-$\Gamma$ which make the nodal line.}
  \end{figure}

\begin{figure}
    \includegraphics[width=3in]{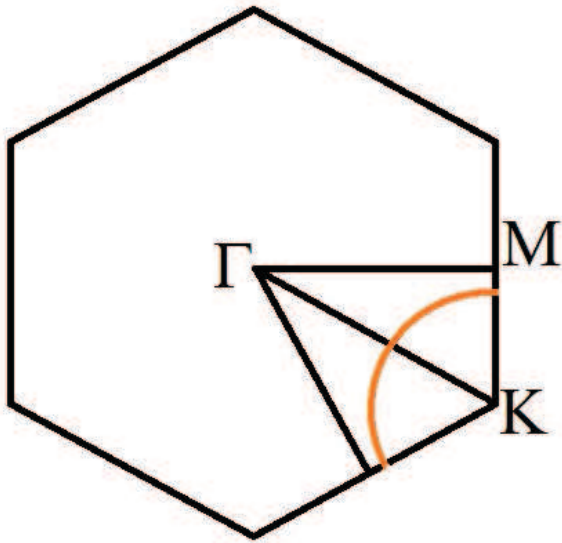}
    \caption{\label{fig:3}(colour online) Projection of the nodal line into the k$_{x}$-k$_{y}$ plane. The nodal is centered at K point, so the other part cross the boundary of the Brillouin Zone.}
    \end{figure}

     Here, we discuss the obtained results of the electronic structure, surface states, and topological Z$_{2}$ invariant  of ZrRuP compound. The calculated band structure  without SOC along high symmetric lines in the Brillouin zone is shown in  Fig.\ref{fig:2} and is similar to other reported resut\cite{Izumi-2002}. The band structure has  high dispersive bands that cross the Fermi level. There are two crossings located along M-K and K-$\Gamma$ high symmetry lines. These crossings make a close nodal, that originates from the  contact of the conduction band and the  valence bands  located above the Fermi level. We perform the calculation of three-dimensional band structure and identify the crosses that make a nodal line. Then, we project this into the k$_{x}$-k$_{y}$ plane and the result is presented in Fig.\ref{fig:3}. The nodal line is centered at K points, and other part  crosses the boundary of the Brillouin zone. The nodal line is protected by the mirror symmetry at k$_{z}$-plane, in the absence of SOC.

\begin{figure}
    \includegraphics[width=3in]{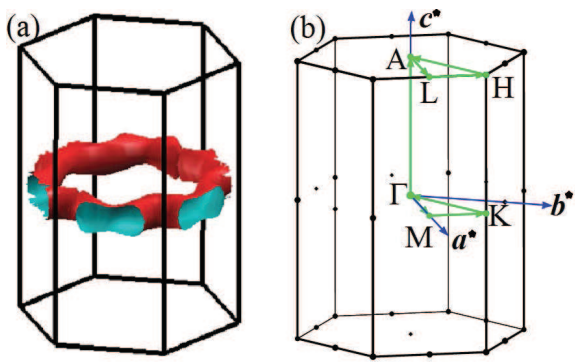}
    \caption{\label{fig:4}(colour online) (a) Band structure of ZrRuP calculated without SOC at energy 0.25 eV. The Fermi surface is formed by connected  torus centered at K point. (b) The Brillouin zone of the hexagonal ZrRuP,  the green line indicates  the path used to calculated the band structure.}
    \end{figure}

   When the Fermi level is shifted by 0.25 eV, it crosses the band that makes the nodal line. This can be achieved by  chemical doping of the system. The corresponding Fermi surface calculated at 0.25 eV  of the bands that make the nodal line is presented in Fig.\ref{fig:4}. The Fermi surface has the form of connected torus centered at the point K. This is consistent with the band structure (Fig.\ref{fig:2}) and the nodal line  (Fig.\ref{fig:3}) because the crossing of the band is located at M-K and K-$\Gamma$, so the nodal line is around K point. One interesting feature of this Fermi surface is that it has open orbits at the boundary of the Brillouin zone which makes this structure different from the previously identified topological materials\cite{Yamakage-2016}.

\begin{figure}
  \includegraphics[width=3in]{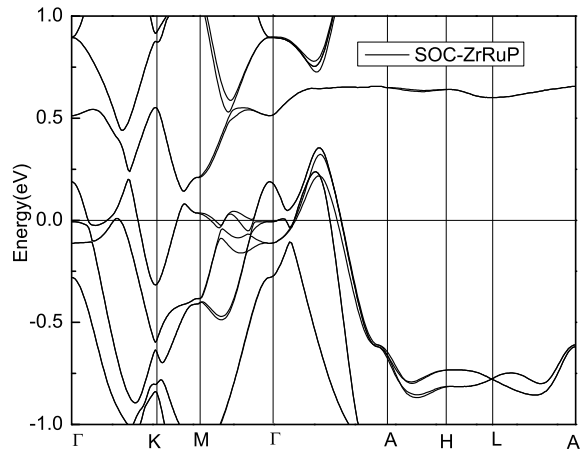}
  \caption{\label{fig:5} Band structure of ZrRuP calculated with SOC along the high symmetry lines on the hexagonal Brillouin zone. In some high symmetry lines the bands split due to the breaking of the inversion symmetry in this structure.}
  \end{figure}

   The inclusion of SOC leads to gap opening across the Brillouin zone in the band structure as   seen in  Fig.\ref{fig:5}.  Since the crystal structure of ZrRuP is noncentrosymmetric, then the degeneracy of the  bands calculated with SOC is lifted. The magnitude of band-gap  is of order 0.1 eV along the nodal line and is not uniform. Therefore, the material does not exhibit the nodal with SOC, and this phenomenon corresponds to a transition from the semimetal phase to adiabatically equivalent to topological insulating phase.

    In order to verify its topological properties, we evaluate the topological invariant Z$_{2}$  \cite{Rui-2011,Moore-2007,Roy-2009}and calculated the corresponding surface states. The  topological invariant number Z$_{2}$ is used to classify the structure about its topological properties, and in three-dimensional is defined by four numbers ($\nu_{0};\nu_{1},\nu_{2},\nu_{3}$). The obtained result of  $Z_{2}$=(0;001), indicates that the material is a weak topological insulator, and the ZrRuP  should exhibit  nontrivial surface states, which will be discussed next.

\begin{figure}
  \includegraphics[width=3in]{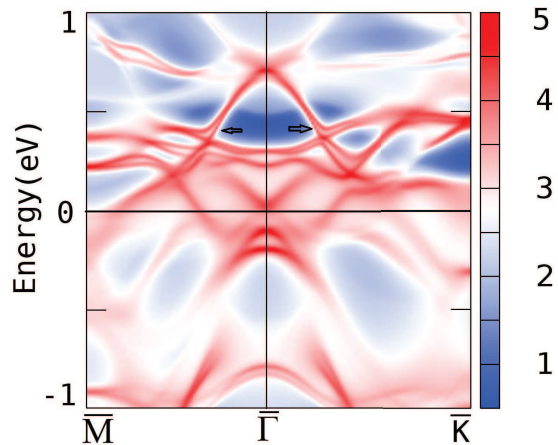}
  \caption{\label{fig:6}(colour online) Surface states  of ZrRuP compound calculated with SOC. There are surface states formed around the band gap region.}
  \end{figure}

   We have constructed the tight binding Hamiltonian of the ZrRuP  which is used for the calculations of the surface states, taking into account the $s$, $p$ and $d$ ( $d_{xy}$, $d_{xz}$, $d_{yz}$, $d_{x^{2}-y^{2}}$ and $d_{z^{2}}$) orbital for Zr and Ru and  $p$ ($p_{x}$, $p_{y}$ and $p_{z}$) orbital for  P . The surface states of ZrRuP  are calculated using the constructed tight-binding Hamiltonian obtained in the presence of spin-orbit coupling. The thin film is constructed using 50 slabs along (001) direction, and the obtained result is presented in  Fig.\ref{fig:6}. The topological surface states appear along the gap region.

   The surface states of the ZrRuP structure is a result of properties of band structure  which  are topological nontrivial. The obtained surface states in this compound can be probed by many different experimental techniques such as the angle-resolved photoemission experiments, which has been used successfully to identify the topological properties of HfSiS\cite{Takane-2016}. This material represents a new type of topological material since it shows a superconducting temperature higher than many topological materials and thus can be of great interest in the field of topological material. In addition, the Fermi surface obtained at 0.25 eV exhibits connected torus which is an interesting feature that could be determined in the future experiments. Next, we discuss the electronic structure of ScPd$_{3}$ that crystallizes in the cubic structure.

    \subsection{Electronic structure of ScPd$_{3}$}

  \begin{figure}
  \includegraphics[width=3in]{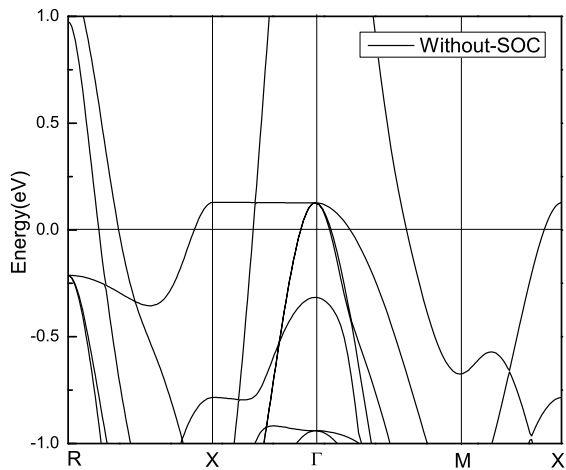}
  \caption{\label{fig:7} Band structure of ScPd$_{3}$ calculated along high symmetry lines in the cubic Brillouin zone. The band structure exhibits different crosses close to Fermi level and make the type-I nodal line above the Fermi level and type-II nodal line below the Fermi level.}
    \end{figure}

   The electronic structures of ScPd$_{3}$ are calculated with and without SOC. Since the structure has the inversion symmetry and time-reversal symmetry the bands should be two-fold degenerate everywhere in the BZ. The Sc and Pd belongs
  to $d$ orbital, so the effect of SOC is non-negligible.Fig.\ref{fig:7} shows the  band structure of ScPd$_{3}$   without SOC, calculated along the high symmetry lines in the first BZ and is consistent with previously reported reslt\cite{Jeong-2006}. From X to $\Gamma$ at the energy of 0.1 eV  there is a crossing which makes the nodal line in the BZ. This nodal line is similar to the nodal line of ZrRuP because it crosses the boundary of the BZ, and also it is located at different energy levels. Another nodal line is formed below the Fermi level along X-$\Gamma$-M at -0.4 eV. This nodal line is characterized by the tilting of the bands and is so called type-II nodal line which is similar to type-II Dirac cone. These nodal lines are protected by the mirror symmetry in the absence of the SOC and are located at the mirror plane k$_{z}$=0.

  \begin{figure}
  \includegraphics[width=3in]{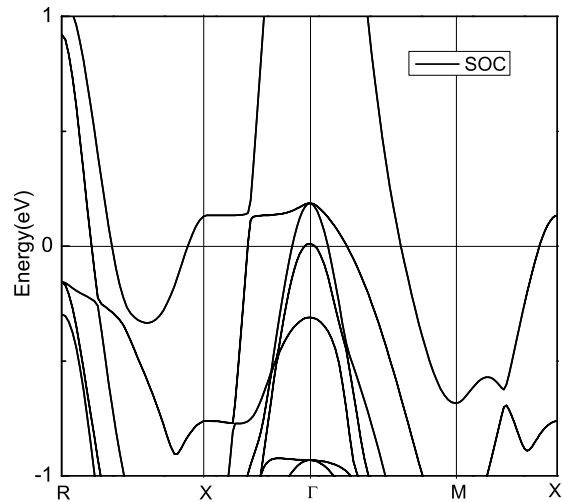}
  \caption{\label{fig:8} Band structure of ScPd$_{3}$ calculated with SOC along high symmetry lines in the cubic Brillouin zone. The band that makes the nodal line at 0.1 eV are gapped out while the nodal line below the Fermi level is not affected by the inclusion of SOC.}
    \end{figure}

\begin{figure}
  \includegraphics[width=3in]{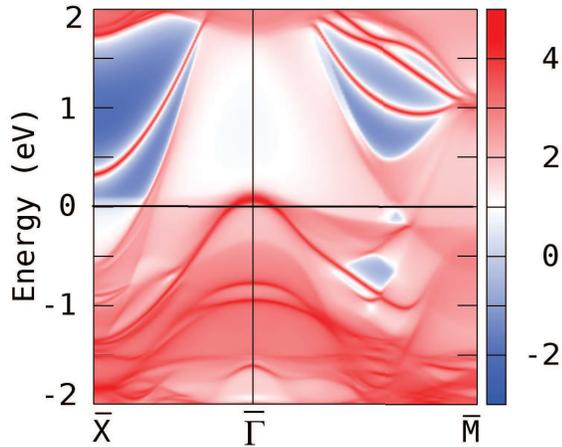}
  \caption{\label{fig:9}(colour online) Surface states of ScPd$_{3}$ calculated with the inclusion of SOC along the (001) diraction. There many surface states that emerge in along the gap region.}
    \end{figure}

   The inclusion of SOC in the band structure calculation leads to gap opening in some nodal lines as shown in Fig.\ref{fig:8} . For instance, the nodal line located at 0.1 eV is fully gapped which leads to adiabatically transition from semimetal to topological insulating phase. Moreover, the nodal line located below the Fermi level is not gapped in the presence of SOC. The calculated topological invariant of ScPd$_{3}$, Z$_{2}$ is (1;000) which demonstrates  that this compound is a strong topological insulator. We also determined the corresponding surface states and the result is presented in Fig.\ref{fig:9}. The ScPd$_{3}$ is characterized by exhibiting nontrivial surface states which can be probed experimentally.  Our calculations indicate that YPd$_{3}$ exhibits similar electronic structure of the ScPd$_{3}$, and is topological nontrivial. We also calculated the electronic structure of ScPd, it is a type-II Dirac material and the Dirac node is located  along M-R high symmetry lines.

  \section{Conclusion}

  We have calculated the electronic structure, and topological properties of ZrRuP and ScPd$_{3}$ by using density functional theory. The band structure of ZrRuP has high dispersive bands that cross the Fermi level which make the close nodal above the Fermi level. The nodal line is protected by the mirror symmetry and is located at the mirror plane $k_{z}=0$. The Fermi level located at 0.25 eV cross the band that makes the nodal line, and the corresponding  geometry of Fermi surface is a connected torus centered at K point. The inclusion of SOC leads to gap opening and the transition to adiabatically equivalent to a topological insulator.  The topological invariant number $Z_{2}$ is (0,001).  The obtained band structure of ScPd$_{3}$ indicates that it exhibits two types of nodal lines located at  $k_{z}=0$. Its corresponding topological invariant is (1;000) which indicates that it is a strong topological insulator. In addition, the structure exhibit nontrivial surface states.  The calculated surface states of ZrRuP and ScPd$_{3}$ appear around the gap region and can be detected by many experimental techniques. We also found that YPd$_{3}$ is a nodal line, while the ScPd is a type-II Dirac semimetal.

  \begin{acknowledgments}

  This work is supported by the Texas Center for Superconductivity at the University of Houston and the Robert A. Welch Foundation (Grant No. E-1146) and the National Natural Science Foundation of China (Grant No. 11574108 ). The numerical calculations were performed at the Center for Advanced Computing and Data at the University of Houston.

  \end{acknowledgments}

\end{document}